\documentclass[preprint,showpacs,nofootinbib,12pt]{iopart}
\usepackage{iopams,setstack}
\usepackage[english]{babel}
\usepackage{ifpdf}
\usepackage{hyperref}
\usepackage{graphicx,xcolor,subfigure}
\usepackage[nodayofweek]{datetime}
\usepackage{textcomp}

\newcommand{\f}{\frac}

\newcommand*{\No}{\textnumero}

\providecommand\ket[1]
{\left\vert \, {#1}\,\right\rangle}
\begin{document}

\title[Mass-jump and mass-bump boundary conditions]{Mass-jump and mass-bump boundary conditions for singular self-adjoint extensions of the Schr\"{o}dinger operator in one dimension}

\author{V.~L. Kulinskii$^{a,b}$, D.~Yu. Panchenko$^{a,c}$}

\address{$^a$ Department of Theoretical Physics and Astronomy, Odessa National University, 2~Dvoryanskaya, Odessa 65082, Ukraine}
\address{$^b$ CREST, North Carolina Central University~\footnote{Fulbright Fellow 2017--2018}}
\address{$^c$ Department of Fundamental Sciences, Odessa Military Academy, 10~Fontanska Road, Odessa 65009, Ukraine}
\eads{\mailto{kulinskij@onu.edu.ua}, \mailto{dpanchenko@onu.edu.ua}}

\begin{abstract}
Physical realizations of non-standard singular self-adjoint extensions for one-dimensional Schr\"{o}dinger operator in terms of the mass-jump are considered. It is shown that corresponding boundary conditions can be realized for the Hamiltonian with the position-dependent effective mass in two qualitatively different profiles of the effective mass inhomogeneity: the mass-jump and the mass-bump. The existence of quantized magnetic flux in a case of the mass-jump is proven by explicit demonstration of the Zeeman-like splitting for states with the opposite projections of angular momentum.
\end{abstract}



\maketitle

\section*{Introduction}

Singular interactions are widely used as theoretical models in quantum physics \cite{qm_deltaberezinfaddeev_dan1961en,book_demkovostrvsk_en,math_deltalbergestexactsolv} and nanotechnology for creation and control of electronic devices\cite{qm_smatrixformicroschem_opertheory1992,qm_zrpmethforcarbonnanotubes_ltp2006,qm_zerorangecarbonnanotubes_jcondmat2007}. The localized structure defects or additional interaction atoms in materials can be modeled by such interactions \cite{qm_zerorangec60_cpl1991}. Even for the simplest case of Schr\"{o}dinger operator of a free particle in one dimension besides the standard case of point-like potential
\begin{equation}\label{freeham}
    \hat{H}_{0} = -\frac{\rmd^2}{\rmd\,x^2}\,,
\end{equation}
there exist other point-like interactions which correspond to self-adjoint extensions of \eref{freeham}. Corresponding boundary conditions were interpreted in terms of inhomogeneous layered materials which can be described with the position-dependent effective mass (see e.g. \cite{qm_deltamassjump_physlett2007,qm_deltamassjump_jphysmath2009}).

Despite the fact that from mathematical point of view the complete mathematical analysis  of the singular point interactions for \eref{freeham} was performed by P.~Kurasov in  \cite{funcan_kurasov_deltadistr_jmathan1996,funcan_deltafinitrank_procmath1998} (see also \cite{funcan_alberverio_singperturb}) the physical interpretation and the possibility of realization of these point-like interactions in real physical systems is far from being clear.
The main result of \cite{funcan_kurasov_deltadistr_jmathan1996} is that the following 4-parameter set of self-adjoint extensions for \eref{freeham}:
\begin{equation}\label{kurasov_d2}
    L_X=-D_{x}^2\left(\, 1 + X_4\,\delta \,\right) + \rmi\,D_{x}\left(\, 2\,X_3\,\delta - \rmi\,X_4\,\delta^{(1)}\right) + X_1\,\delta+(X_2 - \rmi\,X_3)\,\delta^{(1)}
\end{equation}
describe all possible point-like interactions.
Here symbol $D_x$ stands for the derivative in the sense of distributions on the space of functions continuous except at the point of singularity where they have bounded values along with derivatives \cite{funcan_kurasov_deltadistr_jmathan1996,funcan_deltafinitrank_procmath1998}:
\begin{equation}\label{kurasov_delta}
    \delta(\varphi) = \frac{\varphi(+0)+\varphi(-0)}{2}\,,\quad
    \delta^{(1)}(\varphi) = -\frac{\varphi'(+0)+\varphi'(-0)}{2}\,.
\end{equation}
The parameters $X_i$ determine the values of the discontinuities of the wave function and its first derivative. The boundary conditions can be represented in matrix form:
\begin{equation}\label{bc_matrixform}
    \left(
    \begin{array}{c}
        \psi(0+0) \\
        \psi'(0+0)
    \end{array}
    \right)
= M_{X_i}\,
    \left(
    \begin{array}{c}
        \psi(0-0) \\
        \psi'(0-0)
    \end{array}
    \right)\,.
\end{equation}
Physical classification of all these boundary conditions on the basis of gauge symmetry breaking was proposed in \cite{qm_zeropointsymme_physb2015} and summarized in Table~\ref{tab_class}.
\begin{table}[hbt!]
\center
\begin{tabular}{|l|c|}
  \hline
  \hfil BC-matrix &  Physics\\
  \hline
  &\\
    $M_{X_1}=\left(
        \begin{array}{cc}
          1 & 0 \\
          X_1 & 1 \\
        \end{array}
      \right)$ & $\delta$-potential \\
      &\\
  \hline
   &\\
  $M_{X_4}=\left(
        \begin{array}{cc}
          1 & - X_4 \\
          0 & 1 \\
        \end{array}
      \right)$ & mass-bump \\
   &\\
   \hline
   &\\
   $M_{X_3}=\left(
        \begin{array}{cc}
          \frac{2+\rmi\,X_3}{2-\rmi\,X_3} & 0 \\
          0 & \frac{2+\rmi\,X_3}{2-\rmi\,X_3} \\
        \end{array}
      \right)$ & localized magnetic flux\\
   &\\
   \hline
   &\\
   $M_{X_2} =\left(\begin{array}{cc}
                \f{2+X_2}{2-X_2} & 0\\
                0 & \f{2-X_2}{2+X_2}
            \end{array}
        \right)$  & mass-jump \& quantized magnetic flux \\
   &\\
   \hline
\end{tabular}
\caption{Classification of boundary conditions for singular interactions.}\label{tab_class}
\end{table}
Note that although $X_2$-case commonly studied as nonmagnetic (see e.g. \cite{qm_deltamassjump_physlett2007,qm_deltamassjump_jphysmath2009,funcan_kurasov_deltadistr_jmathan1996,qm_massjumphetero_prb1995}), in \cite{qm_zeropointsymme_physb2015} it was shown that in this case the non zeroth quantized magnetic flux is also present. It was also suggested that $X_4$-case corresponds to the presence of mass-jump only and does not include the magnetic field. So this point-like interaction is of potential character like the standard $\delta$-interaction. Both point interactions $X_2$ and $X_4$ are related by the broken dilatation symmetry. The latter means that in these cases we deal with the singularity of the position-dependent mass function.

The aim of this work is to construct physical Hamiltonians with the position dependent mass for $X_2$ and $X_4$ point-like interactions. We also study their difference with respect to the time reversal symmetry caused by the presence of the quantized flux in $X_{2}$-case. This is important from the point of view of the realization of these point interactions in layered sandwich-like materials where the width of transition region is small in comparison with the wave length of a particle.

The structure of the paper is as follows. In Section~\ref{mass-jump_types} we relate the effective-mass Hamiltonian with the regularized form of the Hamiltonian \eref{kurasov_d2} for extensions $X_2$ and $X_4$. We demonstrate that these extensions can be described using the position-dependent effective mass Hamiltonian and correspond to two qualitatively different effective mass profiles. The relation between mathematical parameters $X_{2,4}$ of boundary conditions and the effective-mass singular profile is established. In Section~\ref{sec_loc_mag_X2} we consider point interactions on a circle $S^1$ and show what point interactions have magnetic field via explicit calculation of Zeeman splitting. This way we show that $X_2$-extension belongs to the same ``magnetic`` branch as $X_3$-extension. Namely, $X_2$-extension (so called $\delta^{(1)}$-interaction \cite{funcan_kurasov_deltadistr_jmathan1996,funcan_alberverio_singperturb})  has quantized magnetic flux which was missed in previous works. Possible realizations of these interactions in layered nanomaterials are discussed in Conclusion.

\section{Self-adjoint extensions with position-dependent mass}\label{mass-jump_types}

As has been demonstrated in \cite{funcan_kurasov_deltadistr_jmathan1996} for extensions $X_2$ and $X_4$ the regularized Hamiltonian~\eref{kurasov_d2} can be written as following:
\begin{equation}\label{approx_ham}
    L_{X_2,X_4}^\varepsilon=-D_{x}\left(1 + X_4\,V^{\varepsilon}(x)\,\right)D_{x} + X_2\,V^{\varepsilon\,(1)}(x)\,,
\end{equation}
where
\begin{equation}
    V^{\varepsilon}(x)=\frac{1}{\varepsilon}V\left(\frac{x}{\varepsilon}\right)\,, \quad V\left(\frac{x}{\varepsilon}\right)=\frac{1}{\sqrt{\pi}}\exp\left(-\frac{x^2}{\varepsilon^2}\right)\,, \quad \int\limits_{-\infty}^{+\infty}V(x)\,dx=1\,.
\end{equation}
In the limit $\varepsilon \to 0$ it corresponds to
\begin{equation}\label{kurasov_d2_X4_X2}
    L_{X_2,X_4}=-D_{x}^2\left(\, 1 + X_4\,\delta \,\right) + X_4\,D_{x}\delta^{(1)} +X_2\,\delta^{(1)}\,.
\end{equation}

On the other hand, according to \cite{qm_effectmassem_prb1983,qm_effectmasheter_prb1984,qm_effectmasheter_prb1987,qm_effectmasheter_jpc1988} the Hamiltonian with the position-dependent mass $m(x)$ can be written in the form:
\begin{equation}\label{ham_eff-mass}
    \hat{T}_\alpha=\frac{1}{2}\,m^\alpha(x)\,\hat{p}\,m^{-2\alpha-1}(x)\,\hat{p}\,m^\alpha(x)\,,
\end{equation}
where $\hat{p}=-\rmi\frac{d}{d\,x}$\footnote{$\hbar=1$ in this work.}\,, $m(x)$ is a variable-mass in unit mass $m=1$ and $\alpha$ is some exponent.

\begin{figure}[hbt!]
  \centering
  \includegraphics[scale=0.75]{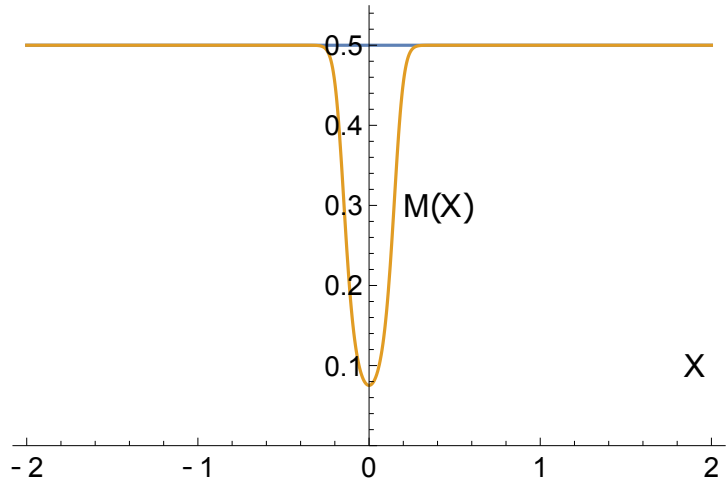}
  \caption{Schematic mass-bump for $X_4$-extension.}\label{fig_massx4}
\end{figure}

As is easy to see that the Hamiltonian~\eref{ham_eff-mass} in the case $\alpha=0$ has the same structure as the Hamiltonian~\eref{approx_ham} in the case $X_2=0$. Thus, the Hamiltonian for extension $X_4$ can be written as:
\begin{equation}
    L_{X_4}^\varepsilon=-\frac{1}{2}D_{x}\frac{1}{m^{\varepsilon}_{X_4}(x)}D_{x}\,,
\end{equation}
where
\begin{equation}\label{mass_X4_eps}
    \frac{1}{m^{\varepsilon}_{X_4}(x)}=2\left(1+X_4\,V^{\varepsilon}(x)\right)
\end{equation}
and $m^{\varepsilon}_{X_4}(x)$ has the form of mass-bump (see Figure~\ref{fig_massx4}).

In addition, in the limit $\varepsilon \to 0$ for the mass-bump case we have the Hamiltonian:
\begin{equation}\label{ham_mass-bump}
    L_{X_4}^\varepsilon=-\frac{1}{2}D_{x}\frac{1}{m_{X_4}}D_{x}
\end{equation}
with inverse mass operator:
\begin{equation}\label{mass_X4}
    \frac{1}{m_{X_4}}=\lim\limits_{\varepsilon \to 0}\,\frac{1}{m^{\varepsilon}_{X_4}}=2\left(1+X_4\,\delta\right)\,.
\end{equation}
The Hamiltonian~\eref{ham_mass-bump} corresponds to the following self-adjoint boundary conditions for \eref{freeham}:
\begin{eqnarray}\label{bc_X4}
    M_{X_4} =
    \left(
        \begin{array}{cc}
            1 & -X_4\\
            0 & 1
        \end{array}
    \right)\,,
\end{eqnarray}
where $X_4$ is defined as:
\begin{equation}\label{X4}
    X_4=\lim\limits_{\varepsilon \to 0}\int\limits_{-\infty}^{\infty}
    \left(\frac{1}{2\,m_{X_4}^{\varepsilon}(x)}-1\right)\,dx\,.
\end{equation}
Naturally, the limiting case $m_{X_4}^{\varepsilon}(x) \to 1/2$ corresponds to $X_4=0$.

In \cite{qm_massjumphetero_prb1995} it was shown that by the coordinate transformation $x\to\eta$ the effective-mass Hamiltonian \eref{ham_eff-mass} can be transformed into common form of a sum of ``kinetic`` and ``potential`` terms:
\begin{equation}\label{ham_eff-mass2}
    \hat{H}_\alpha=\frac{1}{2}\,\hat{p}_\eta^2+\hat{V}_\alpha(\eta)\,,
\end{equation}
where
\[\hat{V}_\alpha(\eta)=\frac{1}{32}(1+4\,\alpha)\left[(1-4\,\alpha)\frac{1}{m}\left(\frac{1}{m}\right)'+4\,\left(\frac{1}{m}\right)''\right]_{\eta}\]
is the effective potential.

Thus the Hamiltonian~\eref{ham_eff-mass} in the case $\alpha=0$ and with the position-dependent mass in the form~\eref{mass_X4_eps} can be represented in standard form:
\begin{equation}
    \hat{H}_{X_4}^\varepsilon=\frac{1}{2}\,\hat{p}^2+
    \frac{X_4}{8}\left[2 V^{\varepsilon}\,''+(X_4 V^{\varepsilon}\,+1) V^{\varepsilon}\,'\right]_{\eta}\,.
\end{equation}
This confirms the results of symmetry analysis of \cite{qm_zeropointsymme_physb2015} that the mass-bump case ($X_4$-extension) belongs to the same class of ``potential`` interactions as the standard $\delta$-potential ($X_1$-extension).

\begin{figure}[th]
  \centering
  \includegraphics[scale=0.75]{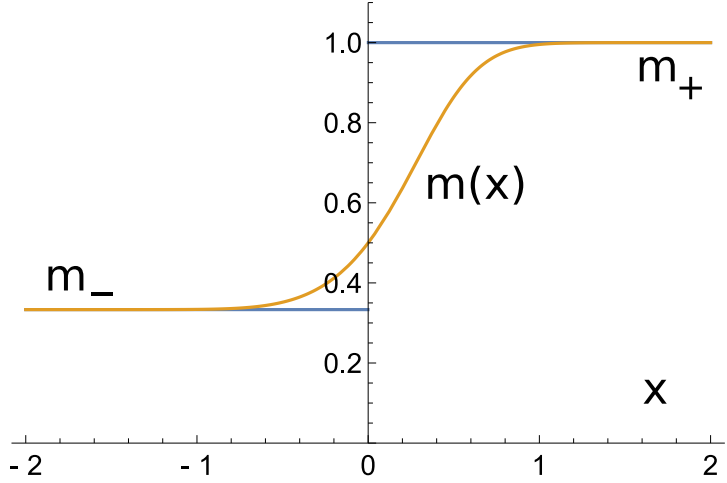}
  \caption{Schematic effective-mass profile of mass-jump for $X_2$-extension.}\label{fig_massx2}
\end{figure}

Note that Hamiltonians \eref{approx_ham} and \eref{ham_eff-mass2} have the same structure if we set $\alpha=1/4$ and $X_4=0$, respectively. Thus, the Hamiltonian for extension $X_2$ can be written as:
\begin{equation}
    L_{X_2}^\varepsilon=-D_{x}^2+
    \frac{1}{2}\left(\frac{1}{m^{\varepsilon}_{X_2}(x)}-
    \frac{1}{m^{\varepsilon}_{X_2}(0)}\right)^{(2)}\,,
\end{equation}
where
\begin{equation}\label{mass_X2_eps}
    \frac{1}{m^{\varepsilon}_{X_2}(x)}=\frac{1}{m^{\varepsilon}_{X_2}(0)}+2\,X_2\,\int\limits_{0}^{x} V^{\varepsilon}(y)\,dy
\end{equation}
so that mass profile has the form of the mass-jump (see Figure~\ref{fig_massx2}).

Now we are able to relate the mathematical parameter $X_2$ with the mass-jump value. According to~\cite{qm_massjumphetero_prb1995}, the boundary conditions for the singular mass-jump are as follows:
\begin{equation}\label{eq_bc_balian}
    \left(
    \begin{array}{c}
        \psi(0+0) \\
        \psi'(0+0)
    \end{array}
    \right)
    = \mathcal{T}\,
    \left(
    \begin{array}{c}
        \psi(0-0) \\
        \psi'(0-0)
     \end{array}
    \right)\,,\quad
    \mathcal{T} =
    \left(
    \begin{array}{cc}
        \mu^{1/4} & 0 \\
        0 & \frac{1}{\mu^{5/4}}
    \end{array}
    \right)\,,
\end{equation}
where $\mu = \frac{m_{-}}{m_{+}}$.

In order to compare~\eref{eq_bc_balian} with that of Kurasov's boundary conditions \eref{bc_matrixform}  we should take into account the appropriate scaling because $X_2$-case corresponds to the dilatation symmetry breaking \cite{qm_zeropointsymme_physb2015,funcan_deltasymm_lettmathphys1998} and therefore the appropriate scale choice on the semiaxes is needed for Kurasov's Hamiltonian. In accordance with the scaling property  $x\to \lambda \,x$ for $x>0$, where $\lambda = \mu^{1/2}$. As $\psi_{0+0}\to \lambda^{-1/2}\,\psi_{0+0}\,,\,\,\psi'_{0+0}\to \lambda^{-3/2}\,\psi'_{0+0}$ then \eref{eq_bc_balian} transforms into:
\begin{equation}\label{eq_bc_balian_scaled}
    \left(
    \begin{array}{c}
        \psi(0+0) \\
        \psi'(0+0)
    \end{array}
    \right) =
    \left(
    \begin{array}{cc}
        \sqrt{\mu} & 0 \\
        0 & \frac{1}{\sqrt{\mu}}
    \end{array}
    \right)\,
    \left(
    \begin{array}{c}
        \psi(0-0) \\
        \psi'(0-0)
    \end{array}
    \right)\,.
\end{equation}
Comparing Kurasov's boundary conditions \eref{bc_matrixform} with \eref{eq_bc_balian_scaled} we obtain the following relation between $\mu$ and $X_2$:
\begin{equation}\label{lambda_balian_kurasov}
    X_2 = \pm 2\,\frac{\sqrt{\mu}-1}{\sqrt{\mu}+1}\,.
\end{equation}
Naturally, the limiting case $\mu \to 1$ leads to $X_2=0$ and if $\mu \to 0$ or $\mu \to \infty$ then $|X_2|\to 2$. Here we do not use of the results of work  \cite{qm_deltamassjump_jphysmath2009} where this $X_2$-case was coupled with $X_1$ interaction. Also we note that from \eref{lambda_balian_kurasov} $X_{2}<2$ and the case $X_{2}>2$ differs only by additional $\pi$-phase. This is indirect hint on he magnetic nature of this singular interaction and we prove it in Section~\ref{sec_loc_mag_X2}.

\begin{figure}
  \centering
  \includegraphics[width=0.7\linewidth]{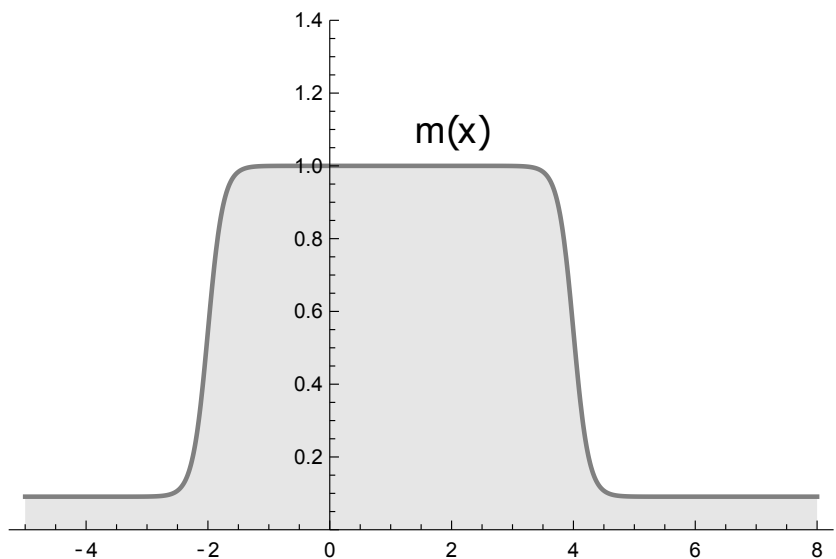}
  \caption{Mass profile for $X_2$-filter.}\label{fig_massfilterx2}
\end{figure}

\begin{figure}[hbt!]
\centering
\includegraphics[width=\linewidth]{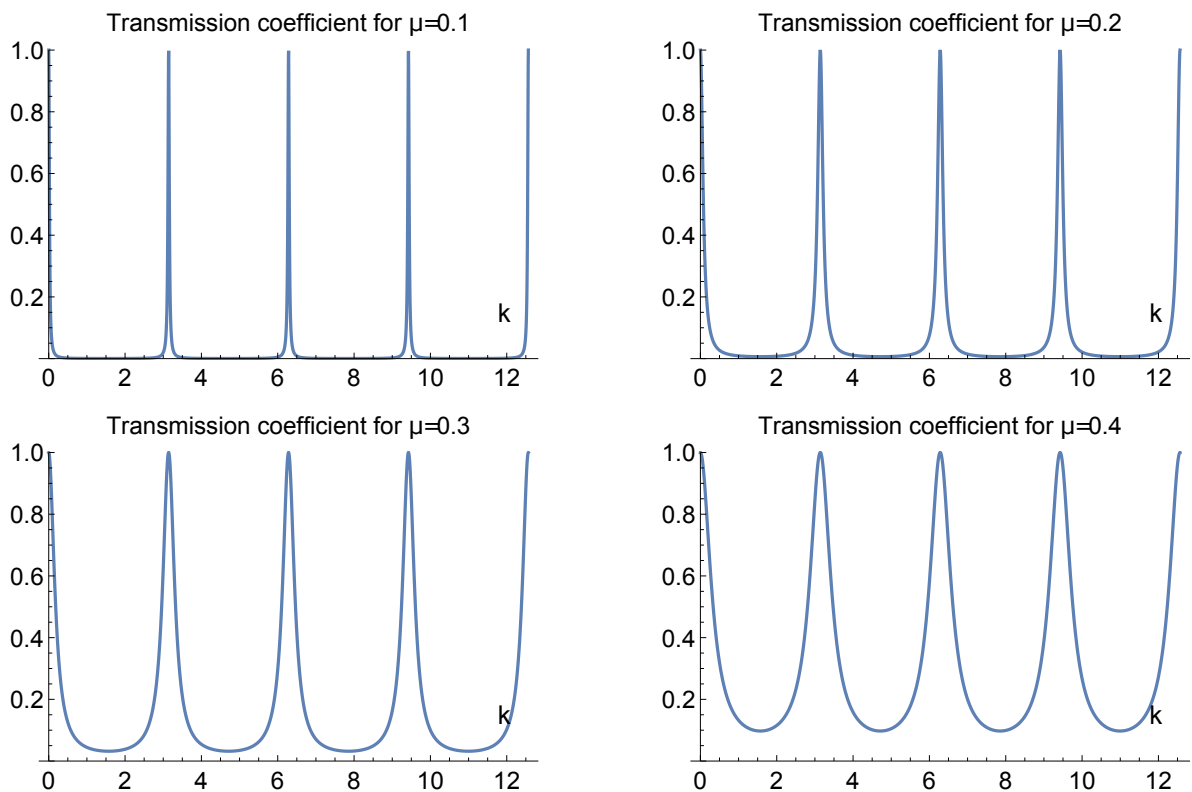}
\caption{Transmission coefficient of $X_2$-filter.}\label{fig_filterx2}
\end{figure}

As an example possible application of such point interaction one can consider the energy filter consisting of two such defects (see Figure~\ref{fig_massfilterx2}). Such filter demonstrates high energetic selectivity as the analysis of transmission coefficient shows (see Figure~\ref{fig_filterx2}).
Despite the fact that $X_2$ point interaction can be described by the Hamiltonian with the position-dependent effective mass in the following section we demonstrate that it also contains a quantized magnetic flux.
\section{Localized quantum magnetic field in $X_2$-extension}\label{sec_loc_mag_X2}
According to symmetry considerations \cite{qm_zeropointsymme_physb2015} $X_2$-case includes non-zeroth localized quantized magnetic flux. From the symmetry point of view $X_2$-extension rather belongs to the same ``magnetic`` branch as the $X_3$-extension associated with the following boundary conditions of localized magnetic flux $\gamma=\frac{\Phi}{2\,\pi}$ which is related to Kurasov's Hamiltonian parameter $X_3$ as following:
\begin{eqnarray}\label{magnet_x3alpha}
 \rme^{\rmi\,2\,\pi\gamma} = \frac{2+\rmi\,X_3}{2-\rmi\,X_3}\,.
\end{eqnarray}
The magnetic nature of $X_3$-extension is obvious and manifest itself also in scattering matrix:
\begin{eqnarray}\label{s_magnetic}
    \hat{S}_{X_3} =
    \left(
        \begin{array}{cc}
          0  & \rme^{\rmi\,2\,\pi\gamma}\\
          \rme^{-\rmi\,2\,\pi\gamma} & 0
        \end{array}
    \right)\,,
\end{eqnarray}
so that $S^{T}\ne S$ which means the time reversal symmetry breaking because of the magnetic field. Sure if $\gamma = n,\,n\in \mathbb{Z}$ we have ``hidden`` non-zeroth flux. This is exactly what happens in $X_2$-extension and we show it below.

\begin{figure}[hbt!]
  \centering
  \includegraphics[scale=0.75]{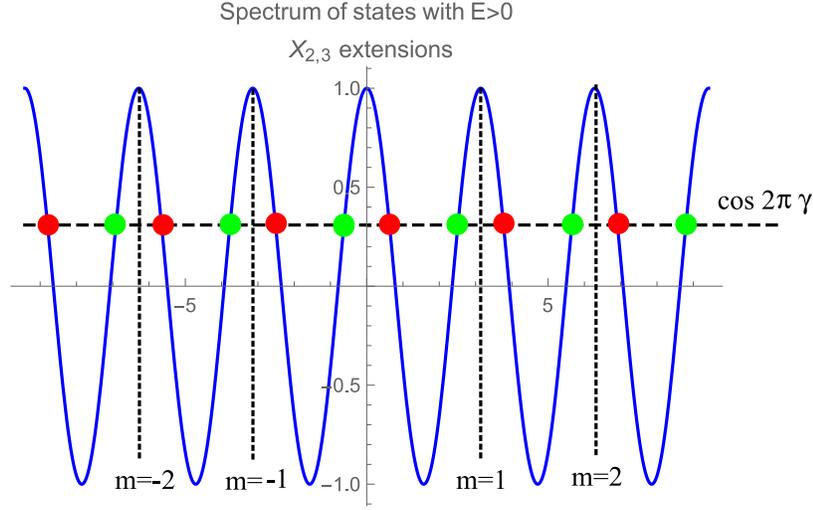}
  \caption{Spectrum of states with $E>0$ on a circle for $X_2,X_3$-extensions. The states of the branch with $\gamma>0$ are colored by red, the ones for $\gamma<0$ branch are green}\label{fig_spectrum_enrgposx2x3}
\end{figure}

\begin{figure}[hbt!]
  \centering
  \includegraphics[scale=0.75]{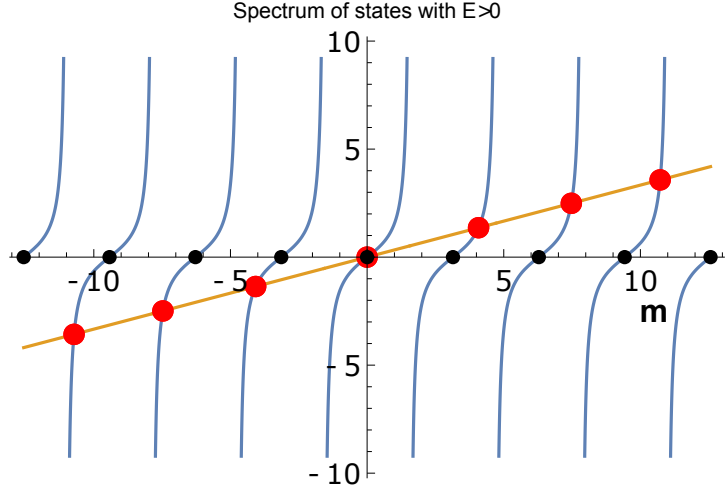}
  \caption{Spectrum of states with $E>0$ on a circle for $X_4$-extension. Black points are standard states, the red ones are emergent due to $X_4$-defect.}\label{fig_spectrum_enrgposx4}
\end{figure}

The magnetic nature of $X_{2}$-extension can be demonstrated by considering it on a circle $S^1$. In such geometry the momentum transforms into angular one and becomes quantized with the magnetic quantum number $\mathfrak{m}$. As is known non-magnetic interactions do not break the degeneration of the energy levels $E_{\mathfrak{m}} = E_{-\mathfrak{m}}$ only magnetic ones do it. Standard solution of the boundary problem on a circle (interval $[-1,1]$ with its ends glued and the singular interaction located at $x=0$) shows that $X_{2,3}$-extensions has identical spectra. For $X_3$-extension the spectrum is given by the equation:
\begin{equation}\label{enrgspectr_x3}
  \cos{2\pi\,\gamma} = \cos{2\,k}\,,\quad E = k^2
\end{equation}
and for $X_2$-extension it is:
\begin{equation}\label{enrgspectr_x2}
  \cos{2\,k} = \frac{2\,\sqrt{\mu}}{1+\mu}\,,\quad E = k^2 \,.
\end{equation}
Obviously \eref{enrgspectr_x3} and \eref{enrgspectr_x2} coincide up to the reparametrization:
\begin{equation}\label{modulo_x2x3}
  \cos{2\pi\,\gamma} = \frac{2\,\sqrt{\mu}}{1+\mu}\,,\quad \mu>0, \,\,|\gamma| \le 1/4
\end{equation}
which gives the relation between the extension parameters $X_2$ and $X_3$ in the appropriate region of parameters $\mu$ and $\gamma$.
But now the states with $m$ and $-m$ are non-degenerate since the states $\ket{\mathfrak{m}+\gamma}$ and $\ket{- \mathfrak{m} +\gamma}$ have different energies $E_{\pm \mathfrak{m}}(\gamma) = (\gamma\pm \mathfrak{m})^2 $, i.e. the splitting due to the magnetic field is $\Delta E = 4\,|\mathfrak{m}|\,\gamma$ (see Figure~\ref{fig_spectrum_enrgposx2x3}). This proves the existence of magnetic field in these cases. The peculiarity of $X_2$-extension is that the ``magnetic`` splitting is caused by the integer magnetic flux but its magnitude is govern by the electrostatic $\mu$-interaction which is attributed to the singular effective-mass profile. The parameter $\mu$ itself is non magnetic in nature by its symmetry though it determines the energetic width of splitting because of the presence of the integer point-like magnetic flux which is unobservable by itself because the boundary conditions matrix $M$ as well as scattering matrix becomes trivial (unit matrix) in such a case.
Also note that in case $X_2 > 2$ additional $\pi$-phase appears although from \eref{lambda_balian_kurasov} it follows that $X_{2}<2$. We treat this as another evidence of the quantized magnetic flux so both integer and half-integer quantized flux are described by singular $X_2$ interaction.
\begin{figure}[hbt!]
  \centering
  \includegraphics[scale=0.75]{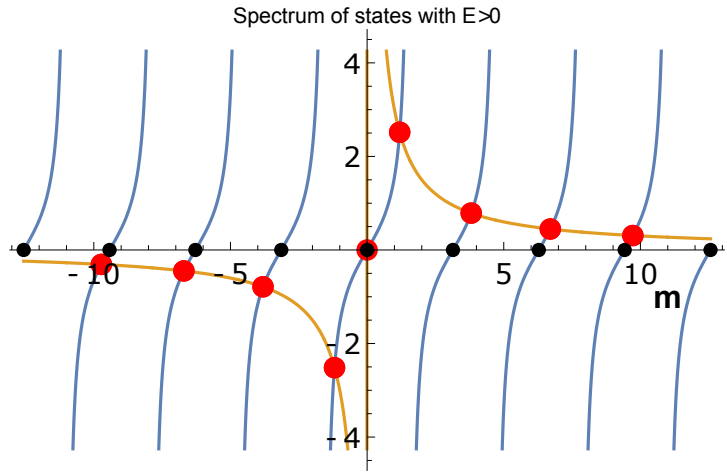}
  \caption{Spectrum of states with $E>0$ on a circle for $X_1$-extension. Black points are standard states, the red ones are emergent due to $X_1$-defect. }\label{fig_spectrum_enrgposx1}
\end{figure}
Now we demonstrate that there is no magnetic splitting for $X_{1,4}$ point interactions on circle. Though the spectra of $X_1$ and  $X_4$ are different the  degeneracy of the energy levels for states $\ket{\pm \mathfrak{m}}$ remains. Indeed, they shift symmetrically with respect to the state with $\ket{m=0}$ (see Figures~\ref{fig_spectrum_enrgposx1}, \ref{fig_spectrum_enrgposx4}).
Namely, the spectrum of states  $E=k^2>0$ for $X_1$-extension is:
\begin{equation}\label{enrgspectr_x1}
\sin{k}\,\left( X_1\,\cos{k} - 2\,k\,\sin{k}\right) = 0 \,\,\Rightarrow\,\,
k^{(0)}_{\mathfrak{m}}=\pi\,\mathfrak{m}
                  k^{(1)}\,\tan{k^{(1)}}= X_1/2
\end{equation}
and the one of $X_4$-extension is:
\begin{equation}\label{enrgspectr_x4}
  \sin{k}\,\left( X_4\,k\,\cos{k}- 2\,\sin{k}\right) = 0 \quad  \Rightarrow \quad k^{(0)}_{\mathfrak{m}} = \pi\, \mathfrak{m}\,,\quad \tan{k^{(1)}} = X_4\,k^{(1)}/2\,.
\end{equation}
Thus we have demonstrated the physical nature of point-like interactions and established the magnetic nature of the point interaction for $X_2$-extension.
\section*{Conclusion}
The paper contains two main results. The first one is that non-standard singular self-adjoint extensions $X_2$, $X_4$ can be described by the position-dependent effective-mass Hamiltonian.
The first case can be realized as the ``mass-bump`` (see Figure~\ref{fig_massx4}). The second one corresponds to the ``mass-jump``(see Figure~\ref{fig_massx2}) with explicit breaking of the dilatation symmetry $x\to \lambda\, x$ \cite{funcan_deltasymm_lettmathphys1998}. In addition, we state that despite this similarity these extensions differ with respect to the time reversal symmetry. A quantized magnetic flux is present in $X_2$-extension, while $X_4$-extension is of pure potential, i.e. ``electrostatic`` nature. Thus, according to the classification of \cite{qm_zeropointsymme_physb2015}, we have two  extensions $X_1$, $X_4$ of the potential nature and two extensions $X_2$, $X_3$ where there is a magnetic field. Possible physical realization of $X_2$-extension can be related with the Josephson junctions and other quasi-one-dimensional heterogeneous structures, where the quantized magnetic flux is localized in the transition layer.
\ack
The authors thank Prof.~Vadim Adamyan for clarifying comments and inspirational insights. The discussions with Prof.~Dmytro Pesin, Prof.~Branislav Vlahovic, Dr.~Maxym Eingorn and Dr.~Vladimir Zavalniuk  are greatly appreciated. This work was completed due to individual (V.K.) Fulbright Research Grant (IIE ID: PS00245791) and with support by MES of Ukraine, grants \No~0115U003208 and \No~018U000202. V.K. is also grateful to Mr.~Konstantin Yun for financial support of the research.

\section*{References}
\providecommand{\newblock}{}

\end{document}